\documentclass[5p, times]{elsarticle}
\usepackage{geometry}
\usepackage{fleqn}
\usepackage{graphicx}
\usepackage{txfonts}
\usepackage{hyperref}
\usepackage{amssymb}
\usepackage{comment}

\usepackage{lineno,hyperref}
\usepackage{graphics}
\usepackage{algorithm}
\usepackage{blindtext}

\usepackage{scrextend}
\usepackage{graphicx}
\usepackage{mathrsfs}
\usepackage{subcaption}
\usepackage{algpseudocode}
\usepackage{algorithmicx}
\usepackage{color}
\usepackage{amsmath}
\usepackage{epsfig}
\usepackage{multirow}
\usepackage{overpic}
\usepackage{colortbl}
\usepackage{blindtext}
\usepackage{amsmath}
\usepackage{siunitx}
\usepackage{pdfpages}
\usepackage{mathrsfs}
\usepackage{upgreek} 
\usepackage{verbatim}
\usepackage[inline]{enumitem}
\usepackage{verbatim}
\usepackage{mathrsfs}

\usepackage{lineno}

\journal{Nuclear instruments and methods in physics research A}

\begin{document}

\begin{frontmatter}

\title{Beam Background Simulation and Experiment at BEPCII}
       \author[First,Second]{H.~C.~Shi\corref{cor1}}
	\ead{shihc@mail.ustc.edu.cn}  
       \author[Third]{B.~Wang\corref{cor2}}
       \ead{wangbin@ihep.ac.cn} 
       \author[Third]{H.~Y.~Shi}
       \author[Third,Fourth]{C.~H.~Yu}
       \author[Third,Fourth]{M.~Y.~Dong}
       \author[First,Second]{H.~P.~Peng}

\cortext[cor1]{Corresponding author1}
\cortext[cor2]{Corresponding author2}

\address[First]{State Key Laboratory of Particle Detection and Electronics, 230026, Hefei, China}
\address[Second]{Department of Modern Physics, University of Science and Technology of China, 230026, Hefei, China}
\address[Third]{Institute of High Energy Physics, 100049, Beijing, China}
\address[Fourth]{University of Chinese Academy of Science, 100049, Beijing, China}

\begin{abstract}
A high-level beam background is a crucial challenge to future upgrades to the BEPCII collider. 
We report on the first separate measurement of main beam background components at BEPCII. 
The separation measurement enables the beam background extrapolation towards the beam parameters assumed for the upgraded BEPCII.
The measured rates of each background component are also compared with a brand-new simulation based on SAD and Geant4.
The discrepancy between experiment and simulation remains 1$\sim$2 magnitudes after calibration with beam lifetime, which should be further investigated.
\end{abstract}

\end{frontmatter}

\begin{sloppypar}
\section{Introduction}

The upgraded Beijing Electron--Positron Collider (BEPCII)~\cite{BEPCII_2007} is the only machine in the world that operates in the $\tau$-charm energy range. 
BEPCII is a double-ring machine, with a design coverage of the beam energy of $1.0$ to $2.1$~GeV (and reaches 2.45~GeV in practical), and a peak luminosity of $1\times10^{33} {\rm cm}^{-2}{\rm s}^{-1}$ at the optimized beam energy of 1.89 GeV. This was achieved at beam currents of $849$~mA $\times$~$852$~mA in April 2016.
The Beijing spectrometer (BESIII) is a general-purpose and the only detector operating in the BEPCII storage ring.
BEPCII has operated successfully for more than 10 years and delivered more than 40 fb$^{-1}$ with different center-of-mass energies. 
These data samples have yielded a number of significant results in physics, including the discovery of a candidate for the tetraquark state Zc(3900)~\cite{Zc3900_2013}.

The age of BEPCII has become an increasingly important problem over time. 
Meanwhile, the physics research in this field now requires more data in the above-mentioned region of energy, especially in region with center-of-mass energy larger than 4~GeV.
Therefore, the Chinese Academy of Sciences (CAS) recently approved a project to upgrade BEPCII, referred to as BEPCII upgrade hereinafter.
BEPCII upgrade aims to extend the beam energy to 2.8~GeV by adding RF cavities for the electron and positron rings, and to improve the peak luminosity by a factor of three ($\mathcal{L}=1.1\times10^{33} cm^{-2}s^{-1}$) at a beam energy of 2.35~GeV by increasing the beam current and slightly compressing the beam size.
The detailed lattice design for the BEPCII upgrade has been provided in Ref.~\cite{Geng_ipac2021}.   
  
The BEPCII upgrade will deliver a higher beam current and a smaller beam size, because of which the beam instability and the beam background become severely owing to collective effects. 
This poses a challenge to the design of storage ring, especially around the region near the interaction point (IP), and the radiation tolerance of detector. 
Therefore, predicting the beam background is necessary for the BEPCII upgrade project as well as projects involving a higher peak luminosity, such as the Super Tau-Charm Facility (STCF) of China. 

The beam background for an electron--positron collider can be classified into two categories, $i.e.$,  beam-related background and luminosity-related background. 
The beam-related background at BEPCII is dominantly induced by the loss of beam particles in the storage ring while the luminosity-related background is generated by beam--beam interaction, and is dominated by physical processes with large cross sections in the annihilation of electrons and positrons.
Physical processes with large cross sections have been studied in general, because of which the luminosity-related background can be predicted with an adequately high accuracy. However, uncertainties persist large for the beam-related background, which is highly reliant on the beam parameters and the storage ring. This requires a careful investigation.

The commission of the SuperKEKB and the Belle~II in the last few years has enabled the examination of beam backgrounds through experiments and Monte Carlo (MC) simulations. 
The beam background on SuperKEKB/Belle~II is simulated by using a framework based on the SAD~\cite{SADWeb} program for particle tracking in the accelerator, and on GEANT4~\cite{Geant4_1,Geant4_2,Geant4_3} program for the interactions and responses of particles in the detector~\cite{Lewis_2019,Natochii_2021}.
In the experiments, the two dominant sources of the beam-related background, the Touschek effect and the beam--gas effect, are separated by changing the beam parameters. The ratios of the background rate between experiments and MC simulations, called the ``data/MC'' ratio, show that the difference between them is within an order of magnitude in most sub-detectors~\cite{Natochii_Snowmass2021}.

In this paper, we present a background experiment at the BEPCII/BESIII with different beam parameters to investigate the beam-related background. 
The experiment was carried out for about 11 hours, and provided essential data to study the Touschek effect and the beam--gas effect.
We also perform comparative results of simulations based on SAD and GEANT4 for the same scenarios and beam parameters as in the experiment. 
The results provide valuable reference for the BEPCII upgrade project and future projects like the STCF.

In Sec.~\ref{sec:BEPCII}, we describe the accelerator and background of BEPCII. 
The background experiment is described in Sec.~\ref{sec:experiment}, including the method, the data analysis, and the results.
The background simulations as well as a comparison of their results with those of the experiment is given in Sec.~\ref{sec:simulation}. 
Finally, we summarize the results of this study and prospects for future background research on the BEPCII upgrades in  Sec.~\ref{sec:summary}. 

\section{Beam-related background at BEPCII}
\label{sec:BEPCII}

The sources of beam-related background in an electron--positron collider with beam energy of the order of gigaelectronvolt (GeV) are as follows: 
(1) the Touschek effect, that is Coulomb scattering among particles in a same bunch;
(2) the beam--gas effect, involving interactions between the particles of the beam and the residual gas molecules in the vacuum chamber, including Bremsstrahlung and Coulomb scattering; 
(3) the injection effect, caused by injected particles with a large amplitude of oscillations due to injection kicker errors; and
(4) synchrotron radiation (SR), SR photons mainly generated by dipole magnets near the IP. 
Of the above, the Touschek effect, beam--gas effect, and injection effect produce unstable particles that are out of their expected orbit and hit the beam pipe.
The unstable particles hitting the inner surface of the beam pipe, called lost particles hereinafter, may directly enter the detector if they lost near the machine--detector interface (MDI) or generate secondary particles to the detector when interacting with the MDI materials. This yields the background in the detector and affects its operation if the frequency is too high.

The main beam parameters of BEPCII are listed in Table~\ref{tab:BEPCIIPara}~\cite{Zhang_IPAC2014}. 
The beam-related background was studied during the design of the BEPCII project through MC simulations~\cite{Jin_2003}, in which  beam--gas effect~\cite{Jin_2004Gas}, the Touschek effect~\cite{Jin_2006Tous}, and synchrotron radiation (SR)~\cite{Zhou_2004SR} were separately investigated to estimate the background rate in BESIII. 
Simulations were performed by using the Decay Turtle program to track the beam particles in the accelerator, and by using GEANT3 to simulate the subsequent interactions and responses of lost particles in the detector.
The background count rate, the total ionizing dose (TID), and the non-ionizing energy loss (NIEL) were used to guarantee the safe operation of the detector.
The result of simulations contribute the design of BEPCII, including the application of several movable collimators in both electron and  positron rings.
The detailed settings of the collimators are listed in Table~\ref{tab:collimator}.
The design of BEPCII is successful as evidenced by its operation over the last decade.
However, it is necessary to re-examine beam background for the BEPCII upgrade project owing to increasing beam current and decreasing beam size.

\begin{table}[htbp]
  \begin{center}
  \footnotesize
  \caption{Main design parameters of BEPCII.}
  \begin{tabular}{l|c}
    \hline \hline
    Parameter &                                Value       \\  \hline
    Optimized energy [GeV]  &            1.89           \\
    Beam current [mA]  &                   910            \\
    $\beta^{*}_{x}/\beta^{*}_{y} [m]$ & 1.0/0.015     \\
    Bunch current [mA]  &                  9.8                \\
    Bunch number         &                  93                 \\
    Emittance [nm.rad]   &                 144                 \\
    Coupling [\%]          &                 1.5              \\
    $\sigma_{z}$ [cm]     &                1.5               \\
    Peak luminosity [$\times 10^{33}{\rm cm}^{-2}{\rm s}^{-1}$] & 1.0   \\
    \hline\hline
  \end{tabular}
  \label{tab:BEPCIIPara}
  \end{center}
\end{table}

\begin{table}[htbp]
  \begin{center}
  \footnotesize
  \caption{Settings of the collimators in BEPCII. All collimators are installed in both electron and positron rings except for OCH04 and ICH04, which are installed only in the electron ring. The movable collimators are marked with an asterisk (*). The negative and positive signs of the positions signify upstream and downstream with respect to the IP, respectively. The $\beta_{x/y}$ values are $\beta_x$ for horizontal and $\beta_y$ for vertical.}
  \begin{tabular}{l|c|c|c|c}
    \hline \hline
    name &  position / m & orientation & aperture / mm & $\beta_{x/y}$ / m \\  \hline
    *OCH02  & -8.2         &  Hor.          & 35    &  25.0 \\
    OCH08  & -27          &  Hor.         &  31      &   12.8 \\
    OCH14  & -46.8       &  Hor.         &  26       &  11.2 \\
    ICH08   & +27.5         &  Hor.         &  30     &  2.8 \\
    \hline
    OCV02  & -7.6          &  Ver.         &  28      &  28.9 \\
    OCV15  & -50.1         &  Ver.         &  15       &  4.6 \\
    OCV16  & -64.8         &  Ver.         &  14      &  11.7  \\
    ICV09   & +28.5         &   Ver.         &  15      &  2.0 \\
    \hline
    *OCH04 & -11             & Hor.          & 35      &  22.9 \\
    *ICH04  &  +11              & Hor.          & 35     &  13.5 \\
    \hline\hline
  \end{tabular}
  \label{tab:collimator}
  \end{center}
\end{table}

We investigated the two dominant beam-related backgrounds in BEPCII, the Touschek effect and beam--gas effect, in different scenarios and with varying the beam parameters.
The experiments were performed in the coasting mode of accelerator, which means the beams are naturally decayed without top-up injection. Therefore, the injection effect produced only a sharp peak in  background rate at the beginning of beam injection, and was thus not considered in this analysis.
The SR background could be ignored because of the long and straight design of lattice before the IP, the relatively low beam energy, and the gold coating with thickness of about 10~${\upmu}$m on the inner surface of beam pipe~\cite{Zhou_2004SR}. 
The beam background rate was measured by the Main Drift Chamber (MDC) of the BESIII.
The MDC is the innermost sub-detector of BESIII surrounding the beam pipe, and is designed to measure the momentum of charged particles.
It consists of 43 layers of sensing wires with radii ranging from $79.0$ to $764.1$~mm, and is divided into an inner part containing eight inner layers and an outer part containing the remaining 35 layers.
During the experiment, the background rate from the MDC was used to analyze the components of background and compare them with those of the MC simulations. 

\section{Beam background experiment}
\label{sec:experiment}
\subsection{Experimental design}
The beam background experiments were performed under different scenarios to better understand the major backgrounds in BEPCII to aid in the design of the BEPCII upgrade project.
When the SR and injection backgrounds are ignored, the total single beam background can be parameterized by: 
\begin{equation}
  \label{eq:BeamReBack}
  O_{\rm SB}=S_{\rm tous}\cdot\frac{I_{t}\cdot I_{b}}{\sigma_{x}\sigma_{y}\sigma_{z}}+S_{\rm gas}\cdot I_{t}\cdot P(I_t)+S_{\rm const}, 
\end{equation}
where $O_{\rm SB}$ is the total single beam background rate that can be represented by count rate of the MDC. 
The first term on the right side of Eq.~\ref{eq:BeamReBack} denotes the Touschek background, which is proportional to beam current ($I_t$) and bunch current ($I_b$), and inversely proportional to beam size ($\sigma_{x}$, $\sigma_{y}$ and $\sigma_{z}$). 
The bunch number ($n_{bunch}$) connects beam current to bunch current: $I_t=I_b\cdot n_{bunch}$.
The second term on the right side of Eq.~\ref{eq:BeamReBack} represents the beam--gas background, which is proportional to $I_t$ and the vacuum pressure. 
In general, the vacuum pressure depends on beam current because the residual gas mainly comes from the interactions between SR photons and the inner wall of beam pipe.
The third term on the right side of Eq.~\ref{eq:BeamReBack} means the background from cosmic rays and electronic noise, which is independent of the beam and can be presented by a constant term.

The Touschek and beam--gas backgrounds can be separated in Eq.~\ref{eq:BeamReBack} by using different bunch currents and sizes, where this affects only the Touschek background. 
We did not consider changing the beam size in the experiment because its measurement is not accurate enough.
The experiment was designed mainly to scan the bunch current with the same total beam current by varying the bunch number. 
The beam size is assumed to be constant during the scan, which will be discussed at the end of Sec.~\ref{sec:dataMC}. 
In the experiment, electron or positron beams were repeatedly injected and then made to decay from a total beam current of $460$~mA with different bunch numbers. 
When the beam current had decayed to $450$~mA, events from the random trigger were selected to measure the count rate.
The scan was repeated for the same target beam current of $450$~mA but with different bunch number varying from 118 to 56, which corresponded to bunch currents in the range of $3.8$ to $8.0$~mA in each instance of decay. 
Random trigger means that the detector was triggered with a frequency of 60 Hz, and a time window of 2~$\upmu$s was used for MDC measurement. 
The constant background was measured without any beam in the storage ring. 

\subsection{Data Analysis}
\label{sec:DataAnalyze}
\begin{figure}[!htp]
  \centering
  \begin{overpic}[width=0.4\textwidth]{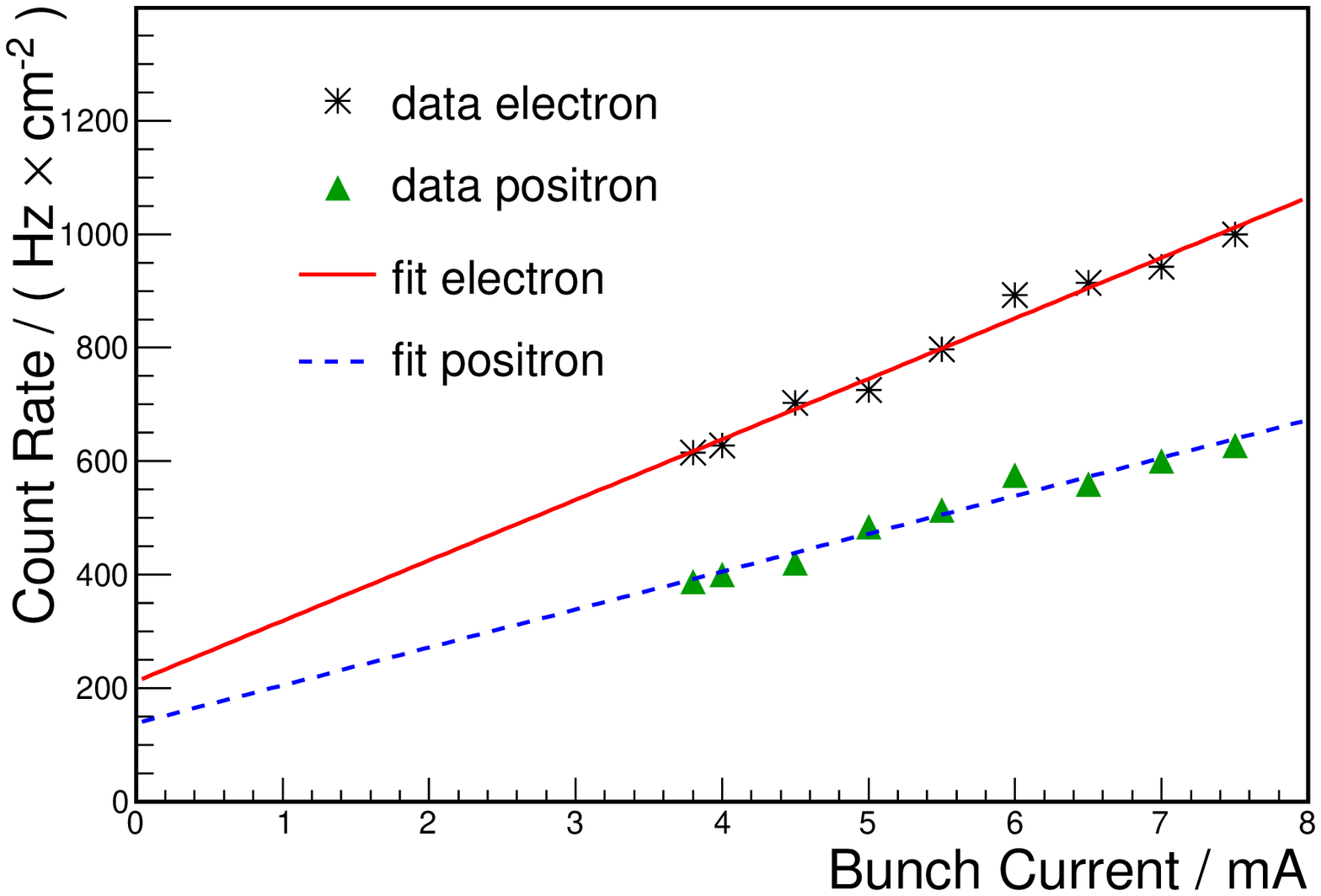}
  \put(80,10){($a$)}
  \end{overpic}
  \begin{overpic}[width=0.4\textwidth]{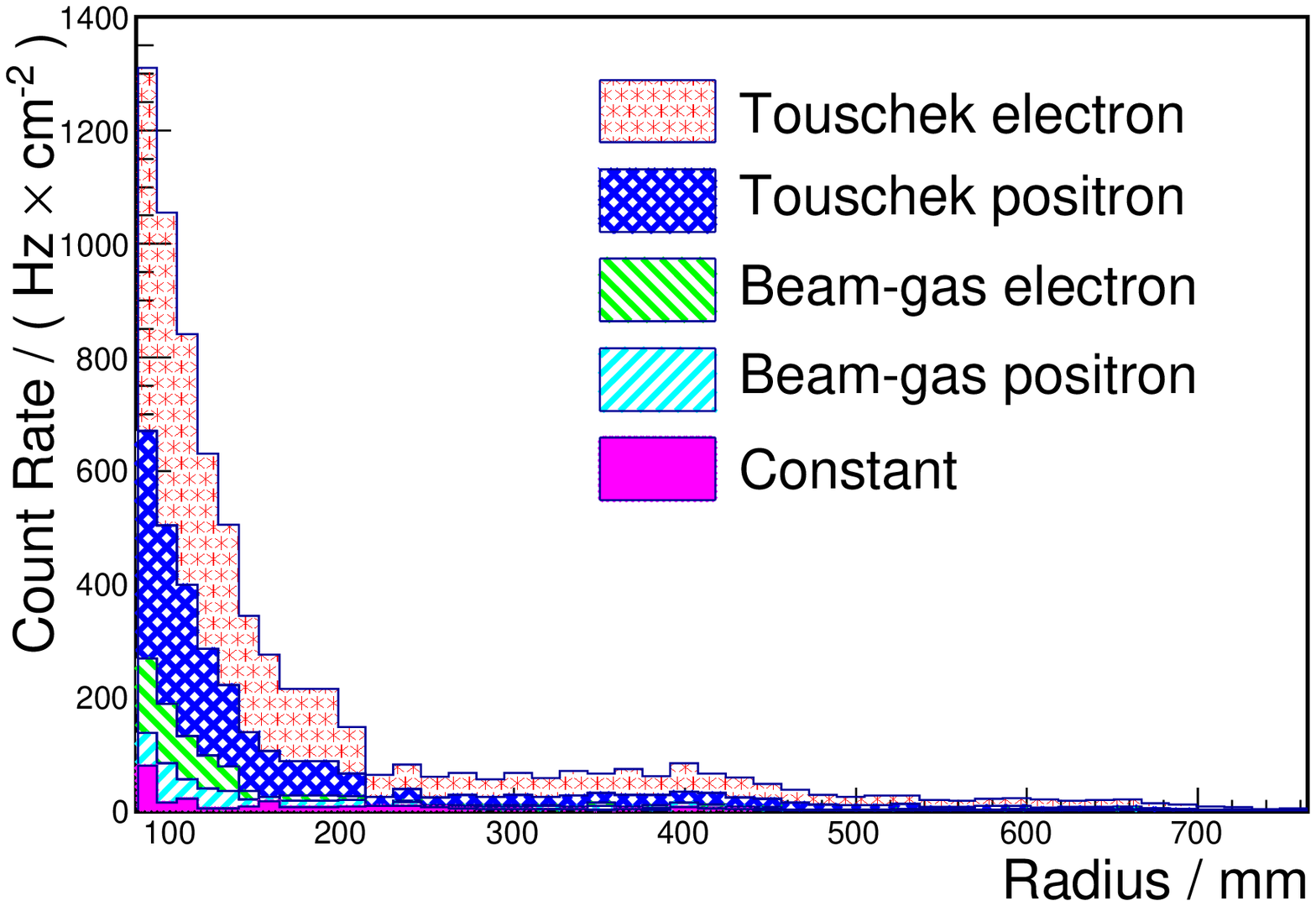}
  \put(80,10){($b$)}
  \end{overpic}
  \caption{
    (a) is the dependence of the count rate in the first layer of the MDC with respect to the bunch current for both electron and positron beams, and the fit on the distribution. The beam current is 450~mA for all points. (b) is the accumulated count rate of separate background sources in all MDC layers when bunch current is 6~mA. 
  }
  \label{fig:FitHitRate}
\end{figure}

When only the bunch current is changed, Eq.~\ref{eq:BeamReBack} can be simplified to $O_{\rm SingleBeam}=k\cdot I_{b}+b$, where $k\cdot I_{b}$ represents the Touschek background and $b$ is the sum of beam--gas and constant backgrounds. 
The constant parameters $k$ and $b$ are obtained by linearly fitting the distribution of the count rate with the bunch current for electron and positron beams, respectively, as shown in Fig.~\ref{fig:FitHitRate} in the first layer of the MDC. 
The beam-related background is then separated at a bunch current of $6$~mA in all MDC layers, as illustrated in Fig.~\ref{fig:FitHitRate}. 
The Touschek background is calculated from the slope in the fit. 
The intercept of the fit includes beam--gas and constant background.
So the beam--gas background is calculated by subtracting the constant background from the intercept. 
Then the separation is applied in all layers of the MDC, as illustrated in Fig.~\ref{fig:FitHitRate} (b).

Fig.~\ref{fig:FitHitRate} (b) shows that the Touschek background is dominant in all layers and the beam--gas background occupies a small portion, especially in the outer layers. 
In some layers, the intercept from the linear fit is lower than the constant background, which means that the beam--gas background is so small that it is covered by fluctuations of the Touschek and constant backgrounds. 
The experimental results are not compared with those from the MC simulations for these layers. 

When the beam current is 450~mA and bunch current is 6~mA, the background fractions are 75.1\%, 15.4\% and 9.5\% for Touschek, beam-gas and constant background, respectively, at the first layer of MDC by adding the count rate of electron and positron beam.

The background fraction obtained from separation measurement enables extrapolation from BEPCII to the BEPCII upgrade project.
When comparing the design parameters of BEPCII and the BEPCII upgrade project from Ref.~\cite{Geng_ipac2021} at optimized beam energy of 2.35GeV, the beam current increase to 2.26 times, the bunch current increased for 1.06 times, the bunch size is compressed to 0.646 times by only considering the emittance in x and y direction and bunch length in z direction. 
Touschek background is increased to 3.71 times and beam-gas background is increased to 2.26 times.
Thus, the total beam-related background in the first layer of MDC in the BEPCII upgrade project will be about 3.2 times higher than that at present according to ratio of background components from this experiment. 

\section{Simulations}
\label{sec:simulation}
The simulations of the beam-related background are based on the framework including generators, SAD to track the particles in the accelerator, and Geant4 to calculate the MDI interactions and the responses of the detector.

The Touschek and beam--gas background are generated through sampling in SAD. 
The BEPCII lattice is sliced to improve the accuracy of generator and that of tracking, and the background particles are generated and tracked in the middle of each slice in SAD. 
The length of sliced lattice is set by striking a balance between the computational power and the accuracy requirements. 
The initial Touschek background particles are generated by sampling two beam particles with a Gaussian distribution and scattering them. The rate of generation is mainly influenced by the energy spread of the scattered particles, the beam size and energy. 
The initial beam--gas background particles are generated by simulating the interactions between the beam particles and the residual gas, including Coulomb scattering and Bremsstrahlung. The rate of generation is proportional to the vacuum pressure. 
The residual gas is assumed as a mixed gas of 80\% hydrogen and 20\% Carbon-Oxide with the same pressure in the total ring.
Parameters used in generator are same with that in experiment, as listed in Table~\ref{tab:SimuPara}.

\begin{table}[htbp]
  \begin{center}
  \footnotesize
  \caption{Parameters used to generate simulation samples for Touschek and beam--gas background.}
  \begin{tabular}{l|c}
    \hline \hline
    Parameter &                    Value       \\  \hline
    Beam energy [GeV]  &      1.849        \\
    Beam current [mA]  &       450            \\
    Bunch number  &             75           \\
    Bunch current [mA]  &       6            \\
    Emittance [nm.rad]   &      144             \\
    Coupling [\%]          &       1.0              \\
    $\sigma_{z}$ [cm]     &      1.5               \\
    Vacuum pressure  [Pa] &  1$\times10^{-7}$  \\
    \hline\hline
  \end{tabular}
  \label{tab:SimuPara}
  \end{center}
\end{table}

\begin{figure}[!htp]
  \centering
  \begin{overpic}[width=0.5\textwidth]{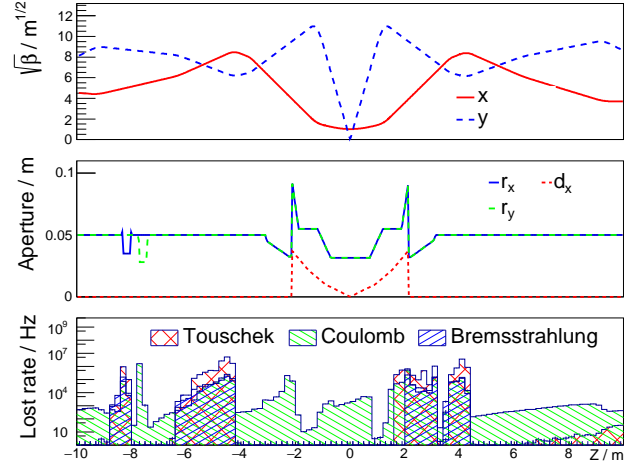}
  \end{overpic}
  \vspace{0.1pt}
  \caption{
Square root of $\beta$ functions, settings of apertures, and the distributions of lost particles with respect to the distance from the IP in the beam direction. The top figure shows the square roots of $\beta$ function in horizontal (x) and vertical (y) directions. The middle figure shows the apertures in x ($r_x$) and y ($r_y$) directions as well as the orbital offset with respect to the center of the beam pipe in horizontal direction ($d_x$). The bottom figure shows the distributions of lost particles due to the Touschek effect (red) and beam--gas effect (green represents Coulomb scattering and blue represents Bremsstrahlung) . 
}
  \label{fig:LostDistribution}
\end{figure}

Once the background particles have been generated, they are tracked slice by slice in SAD. 
The positions of the tracked particles in the plane perpendicular to the beam direction are obtained to determine whether they have reached the inner wall of beam pipe or collimators. 
Collimators with the apertures listed in Table~\ref{tab:collimator} are considered. 
If the positions of particles exceed the aperture within $\pm 10$~m from IP , the lost particles are recorded. 
The bottom plot of Fig.~\ref{fig:LostDistribution} shows the distribution of lost particles with respect to the distance from the IP. 

\begin{figure}[!htp]
  \centering
    \begin{overpic}[width=0.4\textwidth]{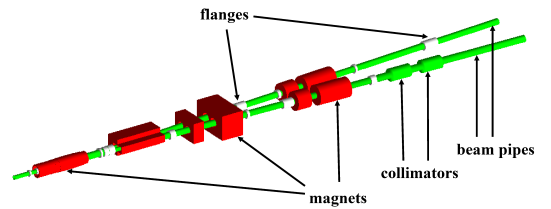}
    \end{overpic}
  \caption{
  Geometry of the MDI elements within 10~m of the IP, including beam pipes, magnets, collimators and flanges. The collimators OCH02 and OCV02 are enveloped by the upstream beam pipe.
  }
  \label{fig:MDI}
\end{figure}

The influence of materials near the IP and the responses of detector are simulated in Geant4. 
We use the Geant4-based software framework of the BESIII Offline Software System (BOSS)~\cite{BOSS}, in which all the sub-detectors have been built and the framework for simulation has been already set. 
The elements of the MDI are added to the framework, as shown in Fig.~\ref{fig:MDI}. 

\begin{figure}[!htp]
  \centering
  \begin{overpic}[width=0.4\textwidth]{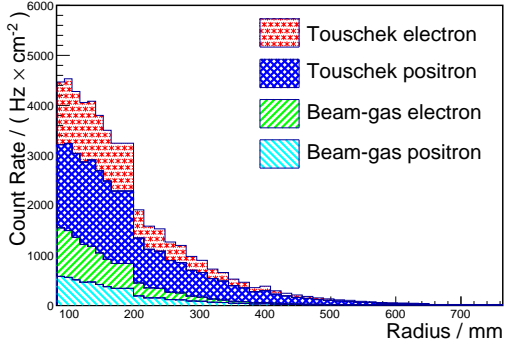}
  \end{overpic}
  \caption{
  The accumulated background count rate of all MDC layers in simulation with beam parameters in Table~\ref{tab:SimuPara}. The beam--gas components are added by Coulomb and Bremsstrahlung scatterings.
}
  \label{fig:hitrate}
\end{figure}

The background count rates in all MDC layers are shown in Fig.~\ref{fig:hitrate}. The Touschek background is dominant. 
The rate of loss and count rate of positrons are both higher than those of electron by about 50\%, mainly owing to two additional collimators in the electron ring in simulation.

\subsection{Comparison between the Experiment and the Simulation}
\label{sec:dataMC}
Before comparing the results of experiment with those of MC simulation, we need to calibrate the background source owing to the incomplete measurement of accelerator parameters. 
Although we set the same parameters in MC simulation with those in experiment, possible difference in beam size and vacuum pressure may lead to difference in generators. 
The beam size is calculated based on emittance and the ideal lattice instead of being measured directly in the experiment. 
But the actual beam size may be affected by blowup effect and fluctuations in $\beta$ function. 
Because the vacuum pressure is measured only near the vacuum pump, it is difficult to estimate the pressure alongside the beam orbit.

\begin{figure}[!htp]
  \centering
  \begin{overpic}[width=0.4\textwidth]{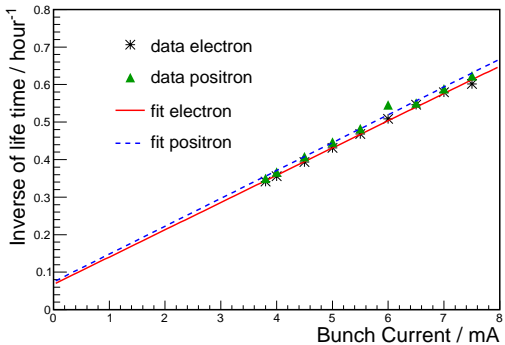}
  \end{overpic}
  \caption{
    Dependence of the inverse of lifetimes of electron and positron beams with respect to the bunch current and the linear fits. The lifetimes of electron and positron beams are close. 
  }
  \label{fig:FitLifeTime}
\end{figure}

We calibrate the generation rate of background by analyzing the beam lifetime, which is parameterized by the following expression assuming that only the bunch current is changed:
\begin{equation}
  \label{eq:LifeTime}
   \frac{1}{\tau}=\frac{1}{\tau_{\rm tous}}+\frac{1}{\tau_{\rm gas}}=k_{\rm tous}\cdot I_{b}+k_{\rm gas}\cdot P, 
\end{equation}
where $I_{b}$ is bunch current and P is the average vacuum pressure along the beam orbit. 
By changing the bunch current with same total beam current, P is assumed to be constant. 
The lifetime of SR effect is negligibly small compared with those of the Touschek and beam--gas effects. 
The lifetime is separated by using the same linear fit as in Sec.~\ref{sec:DataAnalyze}. 
The dependence of the inverse of lifetime with respect to bunch current and the fits are illustrated in Fig.~\ref{fig:FitLifeTime}. 

The lifetimes at bunch current of $6$~mA are $2.29~(2.25)$ h as Touschek lifetime and $14.80~(13.52)$ h as beam--gas lifetime for the electron (positron) beam. 
These lifetimes are calculated in simulation by counting the lost particles in the entire ring with the same parameters as in the experiment, and the results are listed in Table~\ref{tab:lifetime}. 
The ratios of lifetimes in simulation and experiment, referred as scale factors, are used to calibrate the background generation rate in simulation results.
\begin{table}[htbp]
  \begin{center}
  \footnotesize
  \caption{Beam lifetimes $\tau$ in the experiment and the MC simulation, and the scale factors used to calibrate background level.}
  \begin{tabular}{c|c|c|c}
  \hline
                          & $\tau$ (hour) experiment & $\tau$ (hour) simulation & Scale factor \\ \hline
  Touschek $e^-$ &       2.29                     &         11.00                 &   4.8           \\
  Touschek $e^+$ &       2.25                     &         11.81                  &   5.2          \\
  Beam-gas $e^-$ &      14.80                    &         37.26                &   2.5           \\
  Beam-gas $e^+$ &      13.52                    &         37.26                &   2.8           \\     \hline
  \end{tabular}
  \label{tab:lifetime}
  \end{center}
\end{table}

\begin{figure}[!htp]
  \centering
  \begin{overpic}[width=0.4\textwidth]{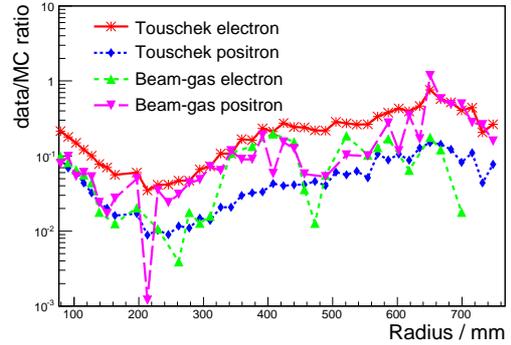}
  \end{overpic}
  \caption{
    Data/MC ratios with respect to the radius of all MDC layers for both electron and positron beams for the Touschek and beam--gas effects. 
  }
  \label{fig:dataMCRatio}
\end{figure}

After multiplying the scale factors on the simulation results, the data/MC ratios for all layers in MDC are illustrated in Fig.~\ref{fig:dataMCRatio}. 
The data/MC ratio shows that the Touschek background in simulation is larger than in the experiment by one to two orders of magnitude. 
Because the generator has already been calibrated with beam lifetime, the difference occurs mainly due to incomplete particle tracking and the interactions of MDI materials in simulation. This needs to be emphasized to improve simulation in the future work.  
Large fluctuations of data/MC ratio for beam--gas background are observed owing to the small count rate and indirect measurements in the experiment. 

We made the assumption that beam size is constant when varying the bunch current. Now we look back to this assumption. 
The beam may expand by blowup effect and Touschek effect may weaken when increasing the bunch current. 
So Touschek(beam--gas) component will be underestimated(overestimated) if beam size is strongly changed by bunch current. 
But Fig.~\ref{fig:FitHitRate}(a) and Fig.~\ref{fig:FitLifeTime} indicate that the correlation of beam size and bunch current is small since beam--gas component is already small and doesn't break the linear relation of background and bunch current. 
Touschek background is still dominant.

\section{Summary}
\label{sec:summary}
In this study, we reported a beam background experiment and simulations at BEPCII. 
The experimental and simulation results both show that the Touschek background is dominant in beam-related background. 
The background fraction obtained from separation measurement makes it possible for background extrapolation from BEPCII to the BEPCII upgrade project as discussed in Sec.~\ref{sec:DataAnalyze}.
The background at the BEPCII upgrade project will be estimated more precisely with separation measurement at beam energy of 2.35~GeV in the future, which is the optimized energy of the BEPCII upgrade project.
Most unstable particles arising due to the Touschek effect oscillated with a large amplitude in horizontal direction. 
Thus, we suggest adding collimators with narrower apertures in horizontal direction. 

A comparison between the results of the experiment and simulations for Touschek background shows a difference of one to two orders of magnitude between them. 
The difference in lifetimes, represented by the scale factors in Table~\ref{tab:lifetime}, is mainly caused by inaccurate estimation of beam size and vacuum pressure on the level of generation. 
The data/MC ratio indicates that the simulations of particle tracking and interactions with MDI materials need to be improved by using a more detailed description of lattice, aperture and MDI geometry modeling. 
As future plan for the BEPCII upgrade project, more background experiments at BEPCII will be carried out to optimize the simulations.

\section*{ACKNOWLEDGMENTS}
This work is supported by the National Science Research and Innovation Fund (NSRF), under Contract No. 11905225, 
and the Xie Jialin Fund No. E25468U210 of the Institute of High Energy Physics of the Chinese Academy of Sciences. 
The authors thank the commissioning group, control group, the mechanical group, and other groups of the BEPCII for constructive discussions and suggestions. We also thank J.~Xun for operational help and guidance for the experiments, and the software group of the BESIII for measurements and essential help in the simulations.

\end{sloppypar}

\end{document}